# Analysis of osteoporotic tissue using combination nonlinear optical imaging

Bryan Semon[1,*], Michael Jaffe[2], Haifeng Wang[3], Lauren Priddy[4], Gombojav Ariunbold[1]

[1]Department of Physics and Astronomy, College of Arts & Sciences, Mississippi State University, 355 Lee Blvd., Mississippi State, MS 39762

[2]Department of Veterinary Clinical Science, College of Veterinary Medicine, Mississippi State University, 240 Wise Center, Mississippi State, MS 39762

[3]Department of Industrial and Systems Engineering, College of Engineering, Mississippi State University, 479-2 Hardy Rd, Mississippi State, MS 39762

[4]Department of Agricultural and Biological Engineering, College of Agriculture and Life Sciences, Mississippi State University, 130 Creelman St, Mississippi State, MS 39762

*bs2302@msstate.edu

**Abstract**

Currently, a large number of stored tissue samples are unavailable for spectroscopic study without the time consuming and destructive process of paraffin removal. Instead, a structurally sensitive technique, sum frequency generation, and a chemically sensitive technique, coherent anti-Stokes Raman scattering enables imaging through the paraffin. This method is demonstrated by imaging collagen in mouse tibia. We introduce a statistical method for separating images by quality and, with the aid of machine learning, distinguish osteoporotic and healthy bone. This method has the potential to verify the results of previous studies and reduce new sample production by allowing retesting results with spectroscopy.

## 1. Introduction

There are approximately one billion formalin-fixed, paraffin-embedded (FFPE) tissue samples currently in storage worldwide.[1,2] This wealth of data is largely inaccessible to spectroscopic techniques as the paraffin itself has a strong signal in many of the same regions as biological material.[3-5] While there are methods to remove the paraffin[5], stain the tissue before embedding[6], or digitally remove the wax contribution[7-9], each has significant drawbacks. Removing the wax is time consuming and, since the preservation is being undone, leaves the sample effectively unable to be stored again.[5] Staining the sample is far more useful for optical imaging as most stains give off an exceptionally strong fluorescence signal, obscuring any other spectroscopic signal.[6] Digitally removing the wax will never perfectly recover the obscured data as well as tending to introduce artifacts.[9]

In this paper we introduce a method of imaging collagen in FFPE samples that is both label-free and nondestructive. This is done by combining two spectroscopic techniques coherent anti-Stokes Raman scattering (CARS) and sum frequency generation (SFG). CARS is a scattering process that generates unique spectra from the vibrotational states of molecules. It produces the same spectral peaks as traditional Raman but with ~$10^6$-fold increase in signal generation.[10] SFG is a multiphoton absorption and reemission process wherein two photons are absorbed, and one photon is emitted. The emitted photon has energy equal to the sum of the incident photons. Since the input beams are spectrally broad for SFG generation, our signal is correspondingly broad. Combining the structural sensitivity of SFG[11] and the chemical sensitivity of CARS [12,13] enabled mapping of both the paraffin and the collagen. This technique also allows for the construction of arbitrary sized mosaic images to be created in an automated way, allowing for large scale features to be captured while maintaining the high resolution of microscopy.

To demonstrate we imaged collagen in the tibias of both healthy mice and mice with alcohol induced osteoporosis, which involves structural changes in the bone such as a decrease in enzymatic crosslinking and an overall decrease in bone density.[14-17] Given the large sample size and relative subtlety of the expected structural changes, a machine learning model was used to classify the samples as machine learning typically excels under these circumstances.[18] A statistical method to separate images by quality was also used, the Spatial Q Test.[19,20] This significantly reduces the size of the data set by automating the removal of images that do not contain collagen. We believe that being able to withdraw additional information out of previously stored samples is invaluable, allowing for the continued reuse of already prepared samples to both verify old studies, as well as to develop new insights based on the original studies using different techniques.

## 2. Methods

### 2.1 Experimental setup

An ultrafast Ytterbium doped fiber pulse laser (Clark-MXR) was used to generate the CARS and SFG signal used here. The initial beam was centered at 1035 nm and had a repetition rate of 1MHz and an initial power of 9.6 W. This beam was passed into a non-colinear optical parametric amplifier (NOPA) to create the three beams necessary: a 1035 nm pump beam, an ~800 nm Stokes beam, and a 517 nm probe beam. The initial beam had a significant spectral width, so the probe beam needed to be passed through a pulse shaper. After the shaper, its spectral width was 10 cm$^{-1}$. The Stokes and pump beam also passed through adjustable neutral density filters to reduce power and avoid degradation of the sample. All three beams were recombined with dichroic mirrors and focused onto the sample through a 10 cm achromatic lens. At the sample, the pump beam had a power of 50 mW, the Stokes 35 mW, and the probe 4 mW. The signal was collected by a long working distance objective lens with a magnification of 50x. The objective lens was infinity-corrected, requiring the use of a tube lens (20 cm focal length) to form an image. Two filters were used in combination to remove any light above 500 nm.

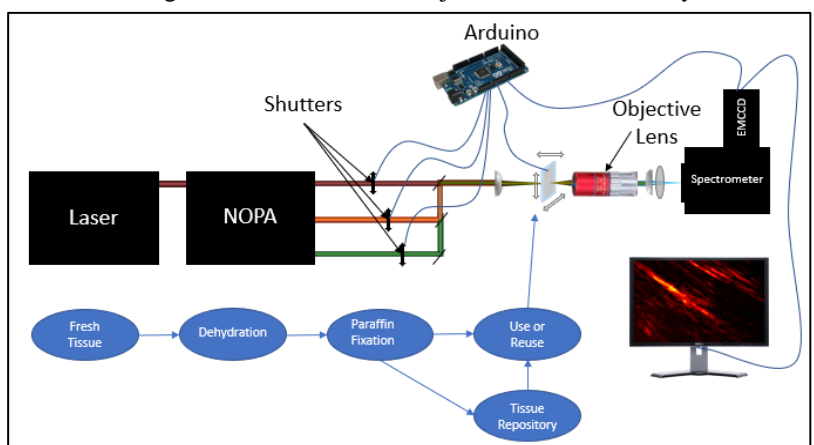

Fig. 1. Simplified diagram of the experimental highlighting the Arduino controlled shutters including the process of obtaining samples.

In order to do near-simultaneous imaging in CARS, SFG, and optical, automatic shutters were needed to block beams. Two beam blocks were 3D printed and motor controls were inserted. The motors were controlled by an Arduino that also communicated with the electron multiplying charge-coupled device (EMCCD) and the sample stage. With this setup, an image was taken with all three beams present (CARS), the probe shutter closed, and another image taken (SFG), and then the pump shutter was closed taking an optical (~800 nm) image. The stage was then moved to a new position and the process repeated. In this way, arbitrarily large mosaic images of the sample in CARS, SFG, and optical were created. Since the images were gathered nearly simultaneously, minimal error (e.g., through beam conditions changing, sample degradation, or misalignment of the sample) was introduced.

*2.3 Sample Preparation*

The animal protocol for this study was approved by the Institutional Animal Care and Use Committee at the University of Southern California (Los Angeles, CA). Alcoholic hepatitis was induced in 8-week-old male C57B/6 mice by feeding a solid Western diet high in cholesterol and saturated fat (HCFD) or regular mouse chow (control) ad libitum for two weeks. Implantation of an intragastric (IG) catheter was performed and IG feeding of ethanol and a high-fat liquid diet (corn oil as 37.1 Cal% [calorie percentage]) at 60% of total daily caloric intake was initiated. Non-alcohol-treated (control) mice were fed a similar high-fat diet. The remaining 40 Cal% was consumed by ad libitum intake of diet high in cholesterol and saturated fat.[21] The amount of alcohol administered to achieve sufficient ethanol intake and blood alcohol levels (BALs) while minimizing the risk of over intoxication was increased in a step-wise progression over 4 weeks. The amount of ethanol fed through the IG catheter increased to 33 g/kg/day over a four week-period from an initial dose of 22.7 g/kg/day.[21] Beginning the second week of the IG feeding, ethanol IG infusion was withdrawn for 5-6 hours and a bolus (3.5-5 g/kg) of ethanol equivalent to that which was withdrawn was given IG, thus mimicking a situation seen in binge drinking in people. The pathology noted in these mice includes a 40-fold to 80-fold increase in osteopontin (OPN) mRNA in the liver. Osteopontin has been associated with the development of bone-related disorders such as osteoporosis. Osteopontin is a phosphoprotein normally secreted by osteoblasts and regulates bone mass by changing local bone remodeling. Abnormal expression of OPN is involved in the development of several metabolic bone disorders, including osteoporosis. [22]

Tibiae were harvested, snap-frozen, and placed in 10% neutral buffered formalin for 48 hours followed by dehydration in 70% and 95% ethanol. Samples were then embedded in paraffin at ~58°C and then cooled at room temperature. The tissue and paraffin were sectioned at 5μm with a microtome warmed to 37°C and the ribbon of tissue/paraffin placed in a warm water bath at 40-45°C. During this process, the paraffin was removed. The tissue samples are then placed on a glass slide and dried at 37°C overnight. The slides were then placed on a warming block at 65°C to melt the wax and bond the tissue to the glass slide. The slide and tissue sample were then stained with hematoxylin and eosin stain.

*2.2 Q score analysis*

Q score provides a quick and statistically valid way to evaluate the relative heterogeneity of an area within an image in comparison with a larger area in the same image.[23,24] An image of quality should have high heterogeneity (areas with high signal and areas with low/no signal). This is especially true of collagen since it comprises long molecules with gaps in between.[24]

The Q score is a measure of the variance of subregions compared to the variance of the area as a whole. The spatial Q score is defined by the following equation[25]:

$$Q = 1 - \frac{\sum_j^M N_{vj} \sigma_{vj}^2}{N_v \sigma_v^2}$$

*( 1 )*

Where $N_{vj}$ is the number of data points (pixels in this case) in the $j^{th}$ subset of v, $\sigma_{vj}^2$ is the variance of the $j^{th}$ subset of v, $N_v$ is the number of data points in v, $\sigma_v^2$ is the variance of v, and

M is the number of subsets. Each image of the mosaic can be divided into subregions and then each image can have its Q score calculated.

Importantly, the spatial Q test allows only binary inputs, so all pixels must be assigned either 1 or 0. To do this, a threshold was picked between 0 and 1 and everything was sorted either to 1 if it was above the threshold or 0 if it was below. The size and therefore number of subregions is also a choice. The size of the subregions should be of a similar order to the size of the structure of interest. In order for this to be statistically valid, these choices should not have an impact on the relative scores of samples.

*2.3 Machine Learning*

Machine learning was performed in MATLAB using the machine learning toolbox. The provided 'OptimizeHyperparameters' function was used to obtain a rough estimate of the best method, learning cycles, leaf size, etc. From there, parameters were manually tuned to minimize the misclassification rate. A deep tree was selected for, as the difference between osteoporotic bone and the control bone was expected to be extremely subtle and computation time was not a significant factor.

The images were processed and analyzed in the following way: First the background levels were subtracted from each image. Since the area in which the laser forms signal is significantly smaller than the full image, pixels from outside the signal forming region were selected as a background. The images are then cropped down to a 100 by 100 pixel region containing only the signal forming region and their maximum brightness is normalized to one. Empty images (i.e., images containing no bone sample) were then excluded via Q score, reducing the number of images from 1620 (45 samples with a 36-by-36 mosaic each) to 669 images that contained SFG signal. This was approximately equally split between control and osteoporotic with 327 of the former and 342 of the later. A variety of image measurements were taken including Q score, $I^2$ score, energy, and entropy (See figure 4 for a full list). Measurements for contrast, correlation, energy, and homogeneity were calculated from an eight-level grey scale co-occurrence matrix. This data was labeled and then randomly split with 30% of the data being withheld for testing while 70% was used for training.

## 3. Results and Discussion

*3.1 SFG vs CARS*

SFG imaging is a powerful tool to image collagen fibers without the paraffin adding noise to the image. Figure 2 shows a comparison of the same area imaged with CARS, SFG, and optical. There was a significant amount of CARS signal exclusively from the paraffin, which had strong Raman lines around 3000 cm$^{-1}$. SFG, however, is more structurally sensitive than chemically sensitive. This means it generates signal only in the collagen fibers, not the paraffin. By comparing the two images, the paraffin was easily distinguishable from collagen. The optical image is used for background subtraction. Since the CARS image had the beam necessary to create an SFG and optical image, the CARS image will always be a stack of all three images, while the SFG will be a stack of the SFG and optical images.

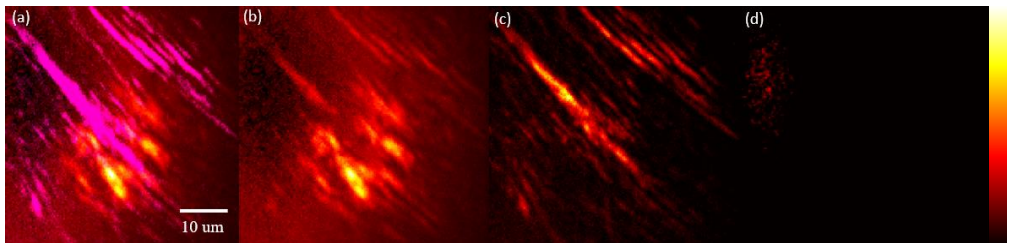

Fig. 2. Image of mouse tibia in CARS with SFG overlay in magenta (a), CARS (b), SFG (c), and optical (d). The striations present in (c) and highlighted in (a) are collagen fibers.

*3.2 Analysis of Q score and Machine Learning Outcomes*

The Q scores for the osteoporotic bone and control bone had a similar average. Visual inspection of the various images agreed with this similarity. Interestingly, the osteoporotic bone had a higher variance in Q scores than the control bone. Looking at this variance for differing thresholds of the Q score shows that this was not a statistical anomaly, and that the Q score results hold, independent of the choice of threshold. Note, the very low and very high thresholds were omitted, as too much noise starts to be included or too much signal is omitted.

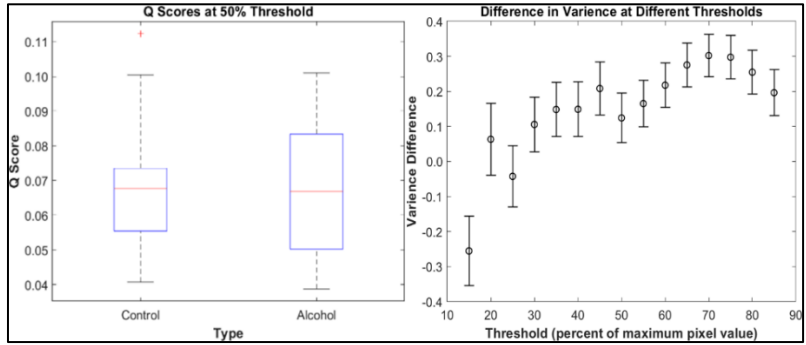

Fig. 3. Box plot of Q scores at a 50% threshold (left). The difference in variance of Q scores for osteoporotic and healthy bone (right). For most thresholds osteoporotic bone has a higher variance in Q scores.

The best performing machine learning model achieved a misclassification rate of 18.9%. It used the AdaboostM1 method with a minimum leaf size of 1, 96 learning cycles, and 65 maximum splits. Figure 4 shows the predictor importance of the best performing ensemble. As can be seen, both the vertical and horizontal corralation were more important than their similar measurements. The Bisque score and entropy were also notably important predictors. The strongest predictor, however, was the Q score. This lends credence to the idea that Q score is a useful measurement of image quality, though on its own was insufficient to distinguish osteoporotic bone from control (healthy) bone.

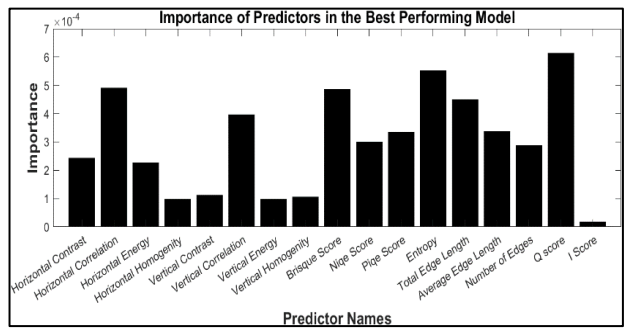

Fig. 4. The relative importance of predictors in the best performing model (AdaBoostM1). The importance was calculated by summing the change in node risk when splitting on a predictor.

Achieving 81% accuracy is lower than we expected for a machine learning model. This we attribute to several key factors. First, sample to sample (i.e., biological) variation; the extent of osteoporosis across individual animals is unknown. Whether or not the structural change due to osteoporosis is uniformly spread across each bone (i.e., within-sample variation) is also unknown. We image a very small area (~1600 µm$^2$) of each bone; therefore, it is possible that some images of ostensibly osteoporotic bone are actually images of relatively healthy portions of the bone. Second, although we have over 600 'good images' of bone, we only worked with forty-five bones in total. Because of the tiling/stitching image method, the twenty-five images from each bone are adjacent to one another, which further exacerbates the issue of limited imaging area.

## 5. Conclusion

SFG imaging allows a method to avoid paraffin contamination in spectral analysis of FFPE tissue, enabling the distinction of the strong Raman signature of the paraffin from the strong SFG signal of the collagen. This novel combinatorial method would allow for more efficient chemical imaging of the large repository of FFPE tissue currently in storage. Imaging in this way produces massive data sets that demand a level of automation. The spatial Q score offers a quick and automated way to evaluate large numbers of images and determine their quality, greatly reducing the size of the data set that must be analyzed in detail. Machine learning also shows promise in separating osteoporotic bone from healthy bone, which may aid in disease diagnoses and general analysis of these large data sets.

## Acknowledgements

We would like to thank Dr. Hidekazu Tsukamoto at the Keck School of Medicine at the University of Southern California for his help in preparing the mice.

## Conflict of Interest Statement

The authors have no relevant conflicts of interest to disclose.